# Evidence of Epitaxial Growth of Molecular Layers of Dissolved Gas at a Hydrophobic/Water Interface

*Ing-Shouh Hwang*[*], *Chih-Wen Yang, and Yi-Hsien Lu*

Institute of Physics, Academia Sinica, Nankang, Taipei, Taiwan

ABSTRACT: The non-wetting phenomena of water on certain solid surfaces have been under intensive study for decades, but the nature of hydrophobic/water interfaces remains controversial. Here a water/graphite interface is investigated with high-sensitivity atomic force microscopy. We show evidence of nucleation and growth of an epitaxial monolayer on the graphite surface, probably caused by adsorption of nitrogen molecules dissolved in water. The subsequent adsorption process resembles the layer-plus-island, or Stranski-Krastanov, growth mode in heteroepitaxy. This finding underlines the importance of gas segregation at various water interfaces and may unravel many puzzles, especially the nature and the high stability of so-called nanobubbles at solid/water interfaces and in bulk water. Based on the hydrophobic effect, we propose that gas molecules dissolved in water may aggregate into clusters in bulk water as well as at solid/water interfaces. As a cluster grows above a critical size, it undergoes a transition into a gas bubble, which can explain formation or nucleation of gas bubbles in water.

---

[*]Address correspondence to ishwang@phys.sinica.edu.tw.

The Hydrophobic effect has been found to play an important role in diverse phenomena, such as protein folding, lipid aggregation, chemical self-assembly, etc. Even though there have been many studies using different theoretical and experimental approaches,[1-22] the microscopic details of how water meets a hydrophobic surface remain elusive. In 1980's, experiments using the surface force apparatus (SFA) detected long-range attraction between two hydrophobic solid surfaces.[1,2] Since then, much effort has been devoted to studying the origin of this "hydrophobic interaction". Several possible mechanisms were proposed.[3] A popular explanation of the attraction is coalescence of nanoscale gas bubbles, or nanobubbles, which might be present on each hydrophobic surface.[4] Indeed, many studies using atomic force microscopy (AFM) reported observation of soft dome-like nanostructures (height of 1.5-100 nm and diameter of 40-2000 nm) on different hydrophobic surfaces in water.[5-15] These three-dimensional (3D) structures were reported to be stable for days,[9] and were considered as surface nanobubbles because supersaturation of gases was required for their formation.[5-14] In addition, nanobubbles do not form from degassed water and degassing leads to their removal.

However, this interpretation of gaseous nanobubbles remains controversial. It has been known that the pressure inside a gas bubble can be described by the Young-Laplace equation,

$$P = P_l + 2\gamma/R \qquad (1)$$

where $P$ is the gas pressure inside the bubble, $P_l$ is the liquid pressure (just outside the bubble), $\gamma$ is the surface tension of the interface between liquid and gas, and $R$ is the radius of the bubble. In recent years, gas bubbles of tens of micron could be efficiently generated in water and their shrinkage with time were also observed.[23,24] The shrinkage rate of those microbubbles was found to accelerate as the bubbles became smaller, which could be well described with the Young-Laplace equation. This leads to an important issue about the high stability of the surface nanobubbles. According to equation (1), a gas bubble having a radius smaller than 1 μm should disappear in less than 1 ms due to its large Laplace pressure.[12,25] The disagreement between the theory and AFM measurements on the lifetime of nanobubbles is at least 10 orders of magnitude![11]

More recently, two-dimensional (2D) gas layers with thickness of 0.3-10 nm and width of micron size, micropancakes, were also reported on highly ordered pyrolytic graphite (HOPG) surfaces in air-supersaturated water, as observed by tapping-mode AFM.[13,14] It is in fact difficult to conceive existence of a 2D gas layer with a thickness of only 0.3 nm and of a 3D gas bubble with a height of only 1.5 nm, because these dimensions are significantly smaller than the mean free path (65 nm) and the average intermolecular spacing (3.4 nm) of gas in the ambient air.

Unfortunately, the existence of gaseous nanobubbles at the interfaces between water and hydrophobic solids was not supported by other experimental studies, such as neutron and x-ray reflectivity measurements,[16-19] sum-frequency vibrational spectroscopic studies,[20] and ellipsometric studies.[21,22] A few years ago, a theoretical study based on molecular dynamics simulations predicted that dissolved gas particles could segregate and adsorb as a monolayer on hydrophobic solid surfaces.[26] In this work, we would like to provide the first experimental evidence of gas adsorption on a hydrophobic surface in liquid.

**RESULTS AND DISCUSSION**

In our experiments, HOPG is used as the substrate because it is moderately hydrophobic (contact angle of ~81.5°)[27] and can be easily cleaved to expose a clean and atomically flat surface. We will show that a certain type of gas molecules, probably $N_2$, forms ordered adlayers at the water/HOPG interface even when dissolved gas is below the saturation concentration. When the dissolved gas level is near saturation, a second and third ordered adlayers as well as 3D structures with height of 1 to 10 nm are also seen. The 3D structures appear similar to the surface "nanobubbles" reported in previous AFM studies. This new finding indicates that the surface "nanobubbles" might be a certain type of condensed matters formed through adsorption of gas molecules at the solid/water interfaces. This can explain many puzzles related to surface "nanobubbles", including their high stability. We will propose an explanation why it is energetically favorable for gas molecules to aggregate into a condensed state, rather than to form a gaseous bubble, when the number of gas molecules is smaller than a critical number. This explanation may also be used to understand the nature of "nanobubbles" in bulk water, which share the same stability puzzle as surface "nanobubbles".[28-31]

In one experiment, we inject pre-degassed DI water to a clean HOPG sample and use an AFM (PicoSPM II from Agilent Technologies) to monitor the time evolution on the surface. The AFM is equipped with an open liquid cell and an environmental isolation chamber. The system has been modified for operation with both the tapping mode and the frequency-modulation (FM) modes. After water injection on the graphite sample at t=0 min, the environmental chamber is exposed to air. Figure 1a shows that only flat and clean graphite terraces are seen at t~40 min. At t~ 60 min, small bright patches start to appear on flat terraces (Figure 1b). Subsequent imaging shows nucleation and growth of the patches (Figures 1c and 1d). These initial small patches are very fragile and can be easily perturbed by the scanning tip. The imaging is stopped temporarily at t~105 min.

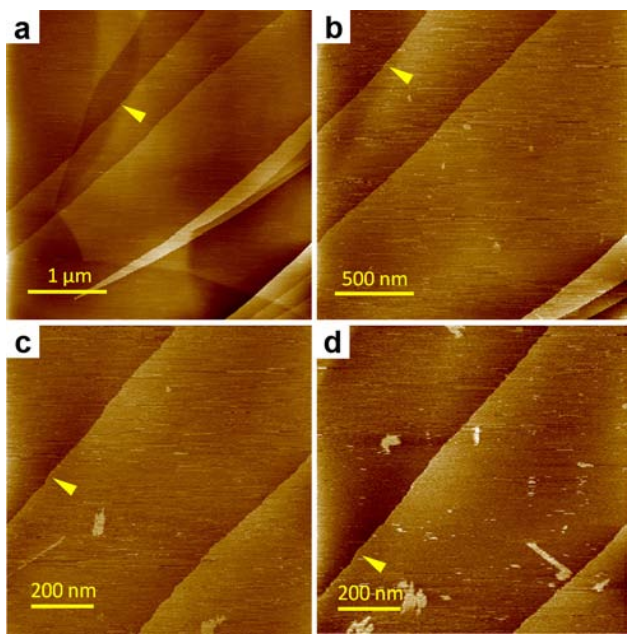

**Figure 1**

When the scanning is resumed at t~210 min, the surface is already ~35% covered by the bright patches (Figure 2a). It clearly indicates that nucleation and growth of the bright patches are spontaneous processes at this interface. Now the bright patches become more stable against the AFM scanning. Figures 2b and 2c show two higher-resolution images of the area, which reveal that the bright patches are composed of domains of a row-like structure. Three different row orientations corresponding to the three-fold symmetry of the HOPG substrate are clearly seen, indicating good registry of the adlayer with the substrate. The row spacing is measured to be 4.2±0.3 nm.

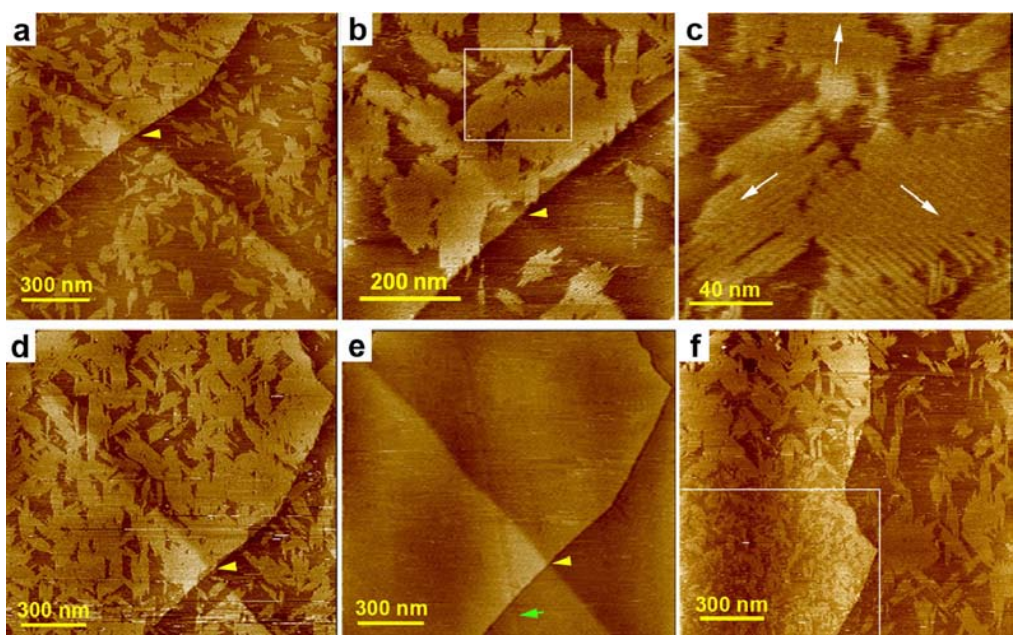

**Figure 2**

We have measured the tip-sample interactions on a bright patch and not on any bright patch by detecting the resonant frequency shift of the AFM cantilever vs the sample displacement (Figure 3). It has been derived that the frequency shift, $\Delta f$, is proportional to the force gradient ($-\partial F_{ts}/\partial z$), where $F_{ts}$ is the interaction force between the tip and the sample.[32,33] Since no frequency shift is detected at large tip-sample separations, we can conclude that there is no detectable long-range electrostatic interaction, suggesting little electrical charge on the bright patches as well as on the graphite surface. At small tip-sample separations (< 1nm), a weak attractive interaction (negative frequency shift) is detected on both cases, which may be related to van der Waals forces between the tip and the sample. On the bright patch, we detect an extra bump before the tip gets in contact with the HOPG surface. This small bump in the resonant frequency indicates a weak repulsive barrier, which may result from the tip's penetrating through the bright patch.

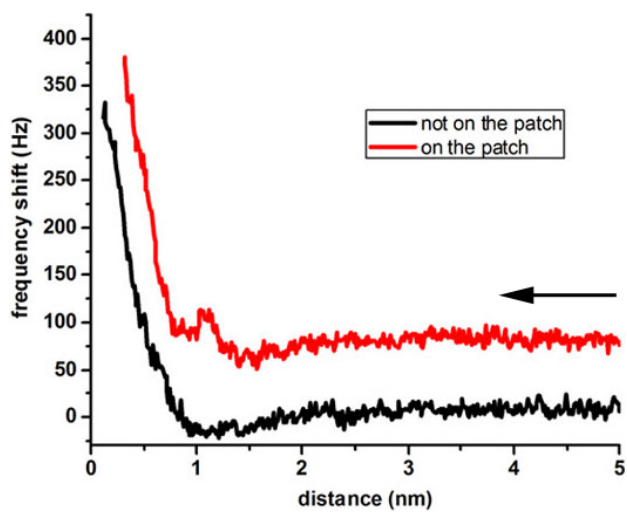

**Figure 3**

At t~380 min, the surface is ~70% covered by the bright patches (Figure 2d). We then switch to the tapping mode and, interestingly, only flat terraces with no bright patches are seen on the same area (Figure 2e). After we switch back to the FM mode, we find that the tapping-mode imaging has strongly perturbed the pattern of the bright patches, as clearly seen in the outlined region in Figure 2f. The area outside the outlined region, which has never been scanned before, exhibits large patches similar to the pattern seen in Figure 2d. This observation explains why previous tapping-mode AFM studies[13,14] did not detect any pattern on HOPG surfaces in DI water or in pre-degassed DI water (see Supporting Information, additional experimental details).

We have observed bright patches on graphite surfaces under pure water in more than 50 rounds of independent experiments conducted over 5 years using four AFM systems located at different locations. Different water containers and pipettes have been used. We have also used DI water taken from three different water purifiers and the bright patches can be consistently detected on HOPG surfaces. Thus it is very unlikely that the bright patches are caused by adsorption of a certain type of contaminants.

Based on previous AFM studies,[5-15] we guess that probably a certain type of dissolved gas molecules adsorb on the HOPG surface to form such an ordered adlayer. Thus, four major atmospheric gases, $N_2$, $O_2$, Ar, and $CO_2$, are examined in the subsequent environmentally control AFM experiments. We first inject pre-degassed DI water on a freshly cleaved HOPG sample in the liquid cell, which is then placed in an environmental isolation chamber and a selected gas is flowed in. Before the gas is flowed in, typically the pre-degassed water has been exposed to air for 2 to 3 min. We find that bright patches comprising domains of the row-like structure will eventually appear for the cases with exposure to $N_2$, $O_2$, and Ar. However, there is a significant difference between $N_2$ and the other two gases. For the experiments with exposure to $N_2$ gas, bright patches often appear at a time less than 60 min after water injection and the growth rate of the patches is higher than the case shown in Figures 1 and 2 (see Supporting Information, Figure S1). Figure 4 shows an AFM image of a HOPG surface in water taken at ~160 min after exposure to $N_2$. One can clearly see domains of the same row-like structure as seen in Figure 2. Now the whole surface is ~70% covered with the bright patches. For the experiments with exposure to $O_2$ or Ar, the growth rates of the bright patches are significantly lower (see Supporting Information, Figure S1). As to the experiments with exposure to $CO_2$, we always detect a strong attraction near the interface (see Supporting Information, Figure S2), which tends to destabilize the AFM imaging. We can thus conclude that the row-like structure is most likely caused by adsorption of dissolved $N_2$ molecules. The reason that the same row-like structure also appears after exposure to $O_2$ and Ar may be due to: the remaining $N_2$ molecules in the pre-degassed water, the initial exposure to air before flowing in the gas, and small air leak of the environmental isolation chamber.

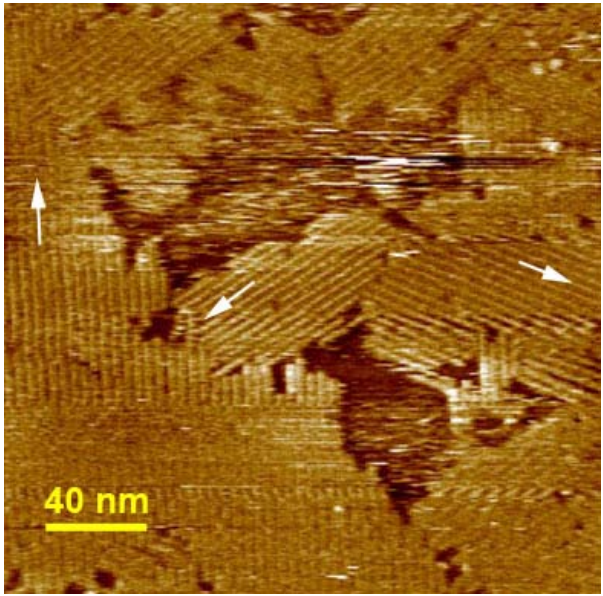

**Figure 4**

Figure 5 shows another AFM study using NanoScope III(a) (from Digital Instrument) located in a different building. This system does not have an environmental chamber, so water is exposed to air after injection. The sample is immersed in a small water droplet, which is trapped between a glass plate and the sample. The water droplet lasts for ~ 2 hr only, so we may add water before the first water droplet is about to evaporate away. On one occasion, about 40% of the surface has been covered with the bright patches at 1.5 hr after a HOPG sample is immersed in a pre-degassed water droplet. Amazingly, we observe formation of higher adlayers and 3D structures on the graphite surface (Figure 5a) after adding a second drop of DI water, which has been exposed to air since the pre-degassed water is taken out from a desiccator. We think the dissolved gas concentration above the graphite sample is close to saturation after adding the second water droplet. Figure 5b shows a high-resolution image with the scan area corresponding to the outlined region in Figure 5a. Now the first adlayer with domains of the row-like structure has covered the entire surface. Three white arrows indicate the three equivalent row orientations related to the HOPG substrate. Interestingly, some areas exhibit the second adlayer with the same row-like structure (one is indicated with"2"). In addition, higher triangular adlayers (three of them are indicated with "3") are observed. On these triangular adlayers, we observe a new row-like structure with a larger row separation of ~5.0 nm. Also, the rows are 30° or 90° rotated relative to those of the first two adlayers. Notice that the surface structure on the triangular adlayers appears somewhat fuzzy, suggesting that there are highly mobile species on the surface. A height profile along the white line is shown in Figure 5c. The second adlayer and the triangular adlayer are about 0.5 nm and 1.0 nm higher than the first adlayer, respectively. On this surface, 3D structures are also present. The ones indicated with "4" and "5" are about 3.0 nm and 1.6 nm higher than the first adlayer, respectively. Many more 3D protrusions can be seen in the larger scan shown in Figure 5a. In fact, the surface morphology resembles the layer-plus-island, or Stranski-Krastanov, growth mode typically seen in heteroepitaxy.[34-36] Figure 5d shows a topographic image taken in a nearby area. A 3D dome-like nanostructure, as indicated with an arrow, has a height of 10 nm and a width of 120 nm, typical dimension of surface "nanobubbles". In addition, there are many other smaller 3D structures that appear to grow on top of the 2D adlayers.

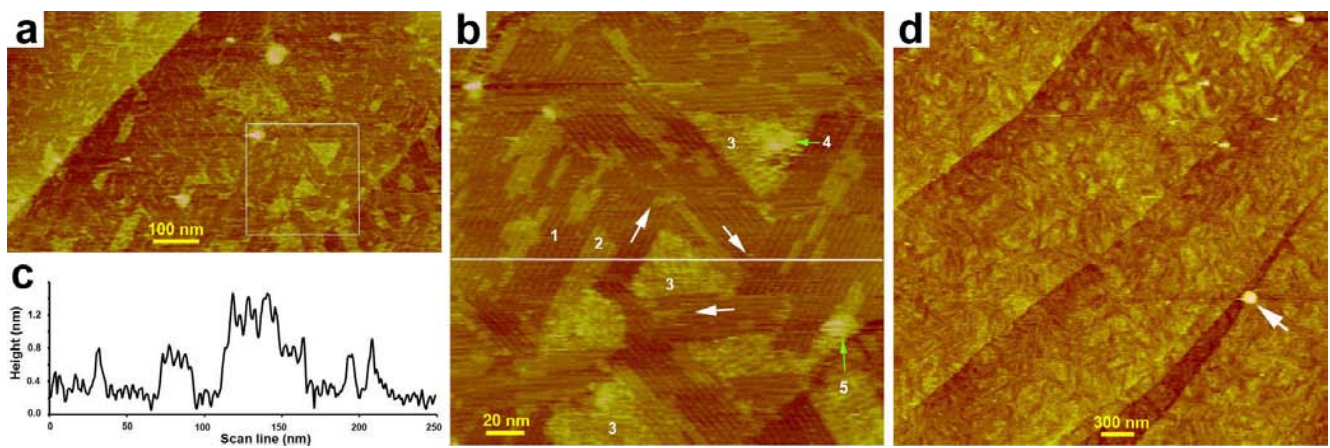

**Figure 5**

A previous theoretical study predicted that dissolved gas molecules might enrich and adsorb at the hydrophobic/liquid interfaces.[26] This prediction was based on molecular dynamics simulations of Lennard-Jones systems. It is not known whether the prediction can be applied to liquid water because the interactions among water molecules are dominated by hydrogen-bonding, rather than by van der Waals interactions. Our AFM observations of ordered adlayers at the HOPG/water interface provide the first evidence that the prediction is valid for water/solid interfaces. In addition, our results indicate that the adsorption can be more than one molecular layer and the adsorbed structures can be ordered. Because the gas adsorption at the interface can occur even when the dissolved gas level is well below saturation, it has further implications for many water interfaces, such as solid/water and oil/water interfaces. If gas molecules dissolved in water can segregate at these interfaces, the interfacial structures and the related properties would be modified. This would also have implications for biological structures, which are often composed of hydrophobic and hydrophilic components.

Our AFM observations further indicate that the 3D nanostructures are condensed matters formed through adsorption of gas molecules at the interface, rather than gaseous "nanobubbles" as widely claimed in the literature. This may be surprising. However, as we will explain below, formation of a cluster with aggregation of gas molecules (a gas aggregate) in water might be energetically favorable over formation of a gas bubble. The interfacial energy per unit area for cluster/water and for cluster/solid wall should be smaller than that for gas/water and for gas/solid wall in a surface bubble, respectively, due to the additional van der Waals attractions at the interfaces. In addition, the total interfacial area is significantly reduced when gas molecules condense from a low-density gas state. These energy gains for the aggregation of gas molecules at the interfaces (scaling as $N^{2/3}$) may outweigh the loss of entropy (scaling as N) when the number of gas molecules in a cluster, N, is smaller than a critical value, $N_c$. If N grows to a size larger than $N_c$, the gas aggregate becomes unstable and may undergo a transition into a low-density gas bubble. This picture is consistent with previous AFM observations of a size limit for surface "nanobubbles", height < 100 nm and diameter < 2 $\mu$m.[10] The stability issue and many other puzzles related to the so-called surface nanobubbles,[9-12] such as contact angles, can also be solved. An immediate application of this finding is high-density gas storage (such as hydrogen, methane, and carbon dioxide) at the ambient temperature and pressure. One may use hydrophobic nanopores or nanostructures, which have high surface areas, for gas adsorption in water.

In recent years, transmission electron microscope (TEM), scanning electron microscope (SEM), and dynamic light scattering (DLS) studies have provided clear evidence that gas molecules can form into nanoparticles in gas-supersaturated bulk water with a lifetime over days or weeks under the ambient conditions.[28-31] Since there are many experimental reports of nanobubbles on solid surfaces, these nanoparticles are also widely accepted as nanobubbles in bulk water. In recent years, there is growing interest in the application of solutions containing "nanobubbles" in diverse research fields, including medical, biological, and environmental applications.[28-31,37] Evidently "nanobubbles" have a nature very different from that of microbubbles, which shrink in size and disappear within a few minutes after their generation.[23,24] The puzzle about the high stability of these "nanobubbles" remains unanswered, as in the case of surface "nanobubbles" observed at solid/water interfaces.

We believe that aggregation of gas molecules into nanoparticles may also occur in supersaturated bulk water. This would explain the observed high stability of "nanobubbles" in solutions for the same reasons as we have described above for surface "nanobubbles". In addition, it is generally known that hydrophobic molecules tend to cluster (or aggregate) in water if their concentration is high enough. Theoretical simulations indicated that hydrophobic clusters must extend beyond a minimum radius (~1 nm) to be long-lived.[38,39] This has been described in terms of the total solvation free energy of *n* hydrophobic particles. If *n* is large enough, the free energy of a cluster is lower than that of the individual solutes, resulting in a driving force for cluster assembly.[39]

Since the major atmospheric gases ($N_2$, $O_2$, and Ar) are nonpolar molecules, they are incapable of forming H-bonds with water molecules and can thus be considered as hydrophobic molecules. Therefore, these gas molecules may well form clusters of nanometer sizes in water. This also suggests a new microscopic picture about how gas molecules are dissolved in water. Some molecules are solvated individually and others form clusters of various sizes, depending on the amount of dissolved gas molecules, the gas type, the conditions (such as the pressure and the temperature of the liquid), and the preparation procedures. Basically, the average cluster size increases with increasing gas concentration.

It is well know that the hydrophobic effect is the tendency of nonpolar substances to aggregate in aqueous solution. The aggregation of gas molecules in water should also belong to this effect. In addition, the segregation of gas molecules at a solid surface, especially a hydrophobic one, should occur more easily than the aggregation of gas molecules in bulk water, just like heterogeneous nucleation versus homogeneous nucleation. An important thing is that there is an upper size limit, $N_c$, for a gas aggregate to be stable. If an aggregate (or a cluster) grows beyond $N_c$, it would transform into a gas bubble. We note that electrical charges might be present at the interfaces between clusters and water[31] and provide significant electrostatic repulsion to prevent large clusters from coalescence into larger clusters that exceed $N_c$. The size limit, $N_c$, is a function of temperature and pressure. An increase in temperature or a reduction in pressure can cause the decrease in $N_c$ due to the favored entropy in the gas phase. Thus, upon a temperature increase or a pressure reduction, large aggregates in a gas-supersaturated solution may exceed the size limit, $N_c$, and undergo transformation into gas bubbles. An example is bubble formation upon opening a bottle of carbonated water. It can also explain bubble formation when supersaturated water is under physical agitations, which force large gas aggregates to collide and coalesce into clusters larger than $N_c$ by overcoming the mutual electrostatic repulsion. Recently, it was reported that small bubbles were formed along the laser beam line in response to a single laser pulse of 9 ns in $CO_2$-supersaturated water and that the threshold laser pulse energy to induce bubble nucleation decreased with the increasing supersaturation level.[40] The pulsed laser beam may provide energy to heat up the $CO_2$-supersaturated water locally, leading to reduction of $N_c$ along the laser beam line. Thus the laser-induced bubble formation can be understood.

**CONCLUSION**

This work demonstrates that high-sensitivity AFM modes can reveal subtle structures at water/solid interfaces. We find that gas molecules dissolved in water may segregate at hydrophobic/water interfaces and form an epitaxial monolayer even when dissolved gas is well below the saturation level. When the gas level is near saturation, higher adlayers and 3D nanostructures are also formed. This finding underlines the importance of considering the segregation of gas molecules at various water interfaces. In addition, our results strongly suggest that the so-called surface nanobbules are probably soft nanostructures formed through aggregation of gas molecules at the solid/water interfaces. Aggregation of gas molecules into clusters may also occur in bulk water. The large clusters might be "nanobbbles" reported in gas-superaturated aqueous solutions. We believe that a cluster is stable when its size is below a critical value. Transition into a gas bubble might occur when the cluster size exceeds the critical value. This picture can explain the formation or nucleation of gas bubbles in water. The findings of this work are of fundamental importance for a wide variety of research fields, including electrochemistry, biology, chemical engineering, environmental engineering, energy, etc. We hope this work can stimulate further theoretical and experimental investigations into the microscopic structures of gas molecules dissolved in liquid and their behaviors in bulk liquid and at solid/liquid interfaces. This would not only better our understanding of a large number of phenomena but also open up many new applications.

**METHODS**

**Materials and Sample Preparation.** Two highly ordered pyrolytic graphite (HOPG) samples are used for the data presented in this work. The one used in PicoSPM II has a size of 20 mm×20 mm (ZYB grade from Momentive) and the other used in Multimode NanoScope III(a) has a size of 10 mm×10 mm (ZYA grade from SPI Supplies). A sample is cleaved right before each AFM experiment. All water used is purified with Milli-Q systems (Millipore Corp., Boston). In preparation of pre-degassed DI water, tubes of purified Milli-Q water (water height ~2 cm) are put inside a desiccator, which is pumped to 0.1-0.2 atm and then sealed without pumping for overnight or longer. The seal is opened immediately before an AFM experiment. The oxygen concentration in the degassed water is measured to be ~10% of the saturation value. The degassed water is extracted with a pipette and then injected into the liquid cell of the AFM. All the experiments are carried out at room temperature.

**Atomic Force Microscopy.** Two AFM systems are used. One is PicoSPM II microscope from Agilent Technologies, which is equipped with an open liquid cell and an environmental isolation chamber. About 400 $\mu$l of DI water is injected into the liquid cell, corresponding to water depth of ~2.2 mm. We have modified the controller for operation with frequency-modulation and a schematic is shown in Figure S3. The other AFM is Multimode NanoScope III(a) from Digital Instrument. The tip holder has been modified with an anti-reflection glass plate that helps the trapping of a water droplet between the sample and the glass plate. The volume of water is ~80 $\mu$l. A schematic is shown in Figure S4. In the AFM images presented in this paper, Si cantilevers (PPP-NCH from Nanosensors) with spring constants 20-40 N/m are used. The resonance frequency and quality factor in water are 120-150 kHz and 8-12, respectively. Right after water injection, significant thermal drift may occur and we often need to wait 30-60 minutes before we can take stable AFM images.

*Conflict of Interest*: he authors declare no competing financial interest.

*Acknowledgment*: We thank support for this work from National Science Council of ROC (NSC96-2628-M-001-010-MY3 and NSC99-2112-M-001-029-MY3) and Academia Sinica

*Supporting Information Available.* Additional experimental details and supporting figures. This material is available free of charge via the Internet at http://pubs.acs.org.

**FIGURE CAPTIONS.**

**Figure 1.** AFM Images of the initial evolution of the water/graphite interface after exposure to air at t=0 min. (a), (b), (c), and (d) are FM-AFM images taken at t= 40, 60, 80, and 100 min, respectively. The height of the bright patches is ~0.45 nm. The yellow arrows indicate a reference position for all images. The resonance frequency of the cantilever is ~142 kHz. The oscillation amplitude is maintained at ~3.6 nm. The frequency shift in the FM mode is typically set between +20 to +40 Hz.

**Figure 2.** AFM images of the evolution of the HOPG surface in water. (a), (b), (c), (d), (e), (f) are images taken at t= 220, 230, 240, 380, 390, and 410 min, respectively. The image shown in (c) is taken in the outlined region in (b). The yellow arrows in (a), (b), (d), (e) indicate a position where a surface step edge (higher height contrast) crosses over a sub-surface step edge (lower height contrast), which might be only 1-3 graphene layers below surface. This position can serve as a marker for comparison of the dynamics of the bright patches between images. The white arrows in (c) indicate the three row orientations of the bright patches, which are parallel with the three equivalent <100> directions of the graphite surface. Only the image shown in (e) is taken with the tapping mode, all others are taken with the FM mode (a higher sensitivity mode). The green arrow in (e) indicates another step edge that is even deeper below the surface than the sub-surface step edge indicated with the yellow arrow. This step edge is hardly seen in images taken with the FM mode, indicating that the FM mode applies a smaller loading force on the surface than the tapping mode. For the FM mode, the operation conditions are the same as in the images shown in Figure. 1. The outlined region in (f) corresponds to the region in the upper-right hand corner in (e). For the tapping mode, the driving frequency is set at the same resonance as the FM mode. The free oscillation amplitude is ~3.6 nm and the amplitude set point is 95%.

**Figure 3.** Approaching curves for the frequency shift vs the sample displacement measured on a bright patch (the upper curve) and not on a bright patch (the lower curve). The upper curve is deliberately shifted upwards for clarity. The frequency shift is zero when the tip is very far away from the sample surface. The tip-sample interactions will cause the frequency shift of the cantilever when the AFM tip is approaching the sample surface.

**Figure 4.** FM-AFM image of a HOPG surface in water after exposure to $N_2$ for 160 min. Three arrows indicate the three row orientations. The tip apex might have small changes during scanning, resulting in different height contrasts of the row-like structure at different scan lines.

**Figure 5.** AFM study of higher adlayers and 3D nanostructures at a water/graphite interface with the phase-modulation (PM) mode. (a) Topographic image of the interface. (b) High-resolution image taken inside the outlined region in (a). First, second, and third adlayers are indicated with "1", "2", and "3", respectively. Three white arrows indicate the three equivalent row orientations of the first adlayer. Two 3D protrusions are also indicated, but there are many other smaller 3D protrusions at this interface. (c) Height profile along the white line in (b). (d) Topographic image of a larger area taken in a nearby

surface. The cantilever working frequency is 142.862 kHz and the phase setpoint is +1.95°. A constant excitation ac signal is used to drive the cantilever and the free oscillation amplitude is ~2 nm.

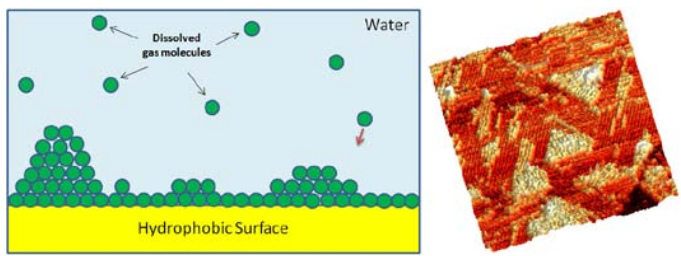

# Supporting Information

# Condensation of Dissolved Gas Molecules at a Hydrophobic/Water Interface

*Ing-Shouh Hwang\*, Chih-Wen Yang, and Yi-Hsien Lu*

Institute of Physics, Academia Sinica, Nankang, Taipei, Taiwan, R.O.C.

Email: ishwang@phys.sinica.edu.tw

**Additional experimental details**

**Observation of bright patches on HOPG Surfaces under DI water.**
The bright patches can be constantly observed with high-sensitivity AFM modes, including the frequency-modulation and the phase-modulation flexural modes presented in this paper. In addition, bright patches can be detected with the torsional modes (frequency-modulation, phase-modulation, and amplitude-modulation) as well as with the PeakForce QNM mode (Bruker AXS). Different types of AFM cantilevers have been used for these different operation modes. We note that the bright patches can also be seen with the tapping mode on rare occasions, even though they usually do not show up in images taken with this mode.

## Supporting Figures

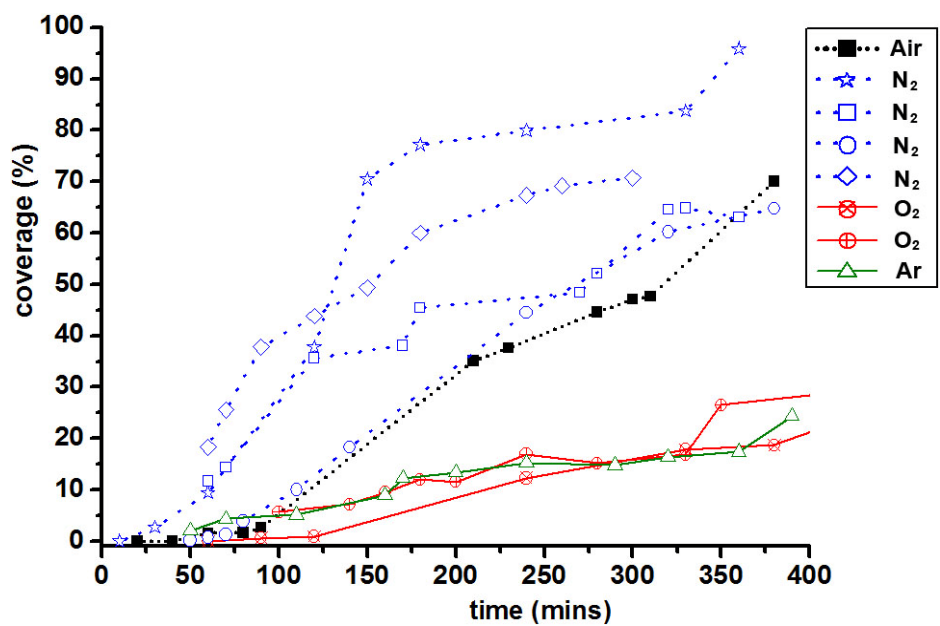

**Figure S1.** Plots of the average coverage of the bright patches versus time after exposure to different gases for HOPG samples immersed in pre-degassed DI water. The plot for exposure to air is drawn based on the data set presented in Figures 1 and 2.

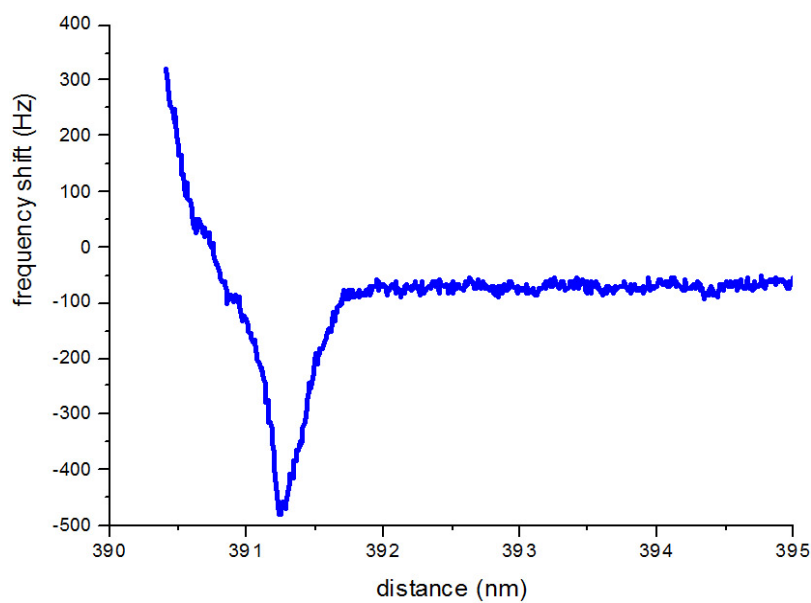

**Figure S2.** Typical approaching curve for the frequency shift versus the sample displacement after exposure to $CO_2$ gas for a HOPG sample immersed in pre-degassed DI water. A strong attraction appears before the tip gets in contact with the substrate.

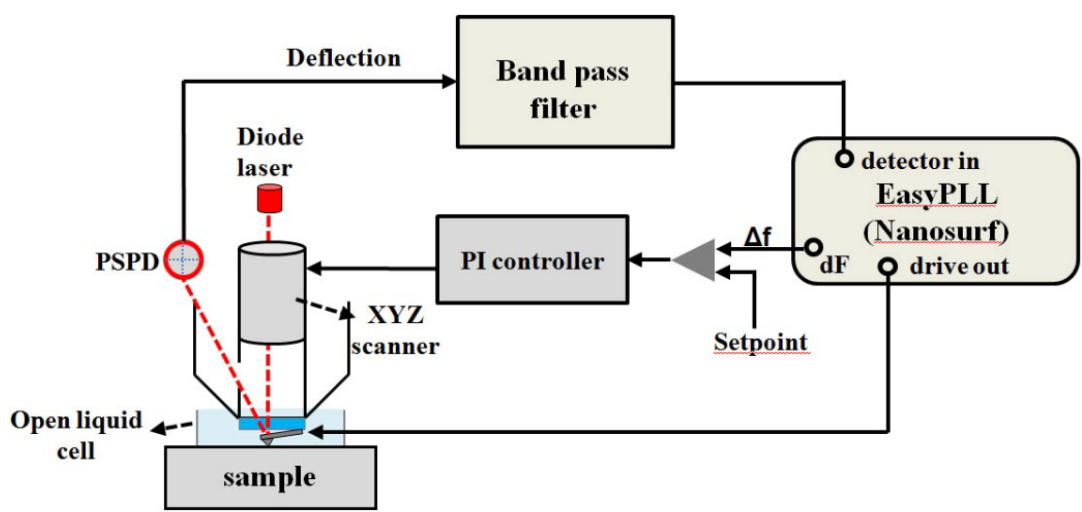

**Figure S3.** Schematic for our modified PicoSPM II microscope. The oscillation of the cantilever is driven with a phase-lock-loop unit (Nanosurf® EasyPLL plus system). We can select the FM/AM detection in the EasyPLL unit. In the AM detection (or the tapping mode), the lock-in mode is selected and the cantilever oscillation amplitude is used as the input signal of the feedback system. In the FM detection, the easyPLL unit is used to track the resonant frequency of the vibrating cantilever and the resonant frequency shift of the cantilever ($\Delta f$) is used as the feedback signal of a PI controller, which outputs a z-axis signal to drive a piezo-scanner tube.

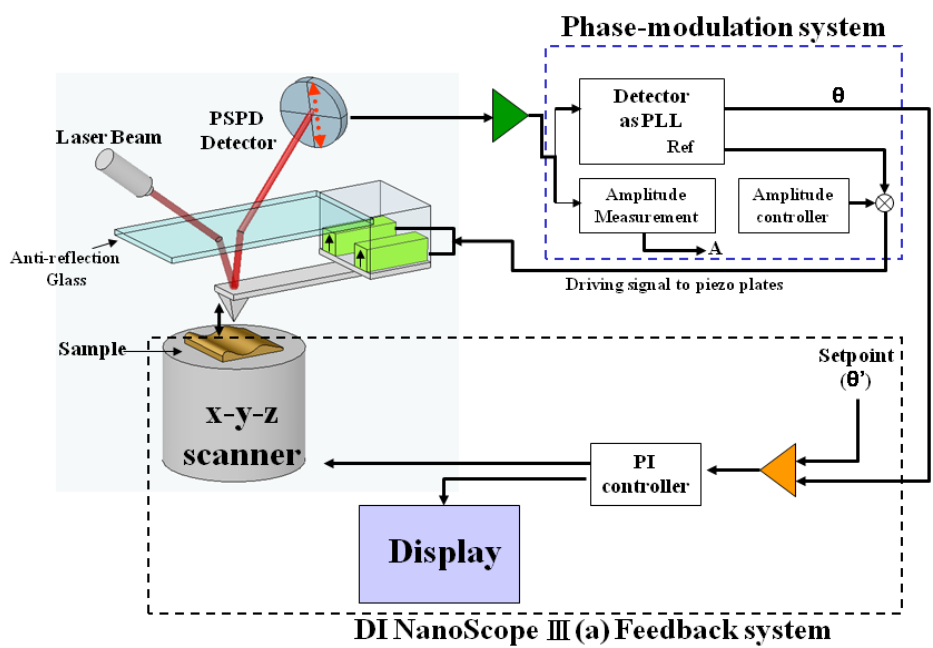

**Figure S4.** Schematic for our modified Multimode NanoScope III(a). The oscillation of the cantilever is driven with a dynamic-modulation system, which is composed of a phase-lock-loop unit (Nanosurf® easyPLL plus system) and a Signal Access Module (Digital Instrument, Santa Barbara, CA). In the phase modulation mode, the lock-in mode is selected and the cantilever oscillation phase is used as the input signal of the feedback system. The tip holder has been modified by adding an anti-reflection glass plate, which helps the trapping of the water droplet and also provides a stable interface for the detection laser beam. For operation in water, a water droplet is introduced between the sample and the glass plate. Note that water drop is not illustrated.